\documentclass{ws-ijmpd}

\usepackage{graphicx}

\begin{document}

\markboth{Gambini and Pullin}
{Consistent discretization adn Regge calculus}

%
\catchline{}{}{}{}{}
%

\title{CONSISTENT DISCRETIZATION AND CANONICAL 
CLASSICAL AND QUANTUM REGGE CALCULUS}

\author{Rodolfo Gambini}
\address{Instituto de F\'{\i}sica, Facultad de Ciencias, 
Universidad
de la Rep\'ublica\\ Igu\'a 4225, CP 11400 Montevideo, Uruguay\\
rgambini@fisica.edu.uy}
\author{Jorge Pullin}
\address{Department of Physics and Astronomy, 
Louisiana State University\\ Baton Rouge,
LA 70803-4001\\
pullin@lsu.edu}

\maketitle

\begin{history}
\received{Day Month Year}
\revised{Day Month Year}
\comby{Managing Editor}
\end{history}

\begin{abstract}
  We apply the ``consistent discretization'' technique to the Regge
  action for (Euclidean and Lorentzian) general relativity in
  arbitrary number of dimensions. The result is a well defined
  canonical theory that is free of constraints and where the dynamics
  is implemented as a canonical transformation.  In the Lorentzian
  case, the framework appears to be naturally free of the ``spikes''
  that plague traditional formulations. It also provides a well
  defined recipe for determining the integration measure for quantum
  Regge calculus.
\end{abstract}


\hfill {\em Dedicated to Rafael Sorkin on his 60th birthday.}

\section{Introduction}
Regge calculus\cite{Re} has been proposed as an approach to classical
and quantum general relativity. It consists in approximating
space-time by a simplicial decomposition. The fundamental variables of
the theory are the lengths of the edges of the simplices. This
approach has been demonstrated in numerical simulations of classical
general relativity and also has inspired attractive ideas for the
quantization of gravity. For instance an extension of this framework
led to the successful quantization of $2+1$ dimensional Euclidean
gravity through the Ponzano--Regge model\cite{PoRe}, which can also
be seen as one of the key motivations for the ``spin-foam'' approaches
to $3+1$ dimensional quantum gravity. There has been quite a bit
of work devoted over the years to Regge calculus, for a recent review
including related formulations see Loll\cite{Lo}, and for a earlier
pedagogical presentations see Misner, Thorne and Wheeler\cite{MTW}.

A canonical formulation for Regge calculus has nevertheless, remained
elusive (for a review see Williams and Tuckey\cite{WiTu}). We have
recently introduced a methodology to treat discrete constrained
theories in a canonical fashion, which has been usually called
``consistent discretizations''.  The purpose of this paper is to show
that this methodology can be successfully applied to Regge calculus
without any need for modifications of the Regge action. The resulting
theory is a proper canonical theory that is consistent, in the sense
that all its equations can be solved simultaneously. As is usually the
case in ``consistent discretizations'' the theory is constraint-free
(although as is usual in Regge calculus there are triangle
inequalities to be satisfied among the variables).  We will see that
the treatment can be applied in both the Euclidean and Lorentzian
case. In the latter case there is an added bonus: in order to have a
well defined canonical structure one naturally eliminates ``spikes''
that have been a problem in Regge formulations in the
past\cite{spikes} at the time of considering the continuum limit. This
is due to the fact that our simplices only have one time-like hinge.
It is therefore not possible to construct simplices with infinitesimal
volume and arbitrary length.  If one lengthens the time-like hinge one
necessarily has to lengthen the space-like hinges and therefore
increase the volume. Therefore one will not see the quantum amplitude
dominated by long simplices of vanishing volume.

\section{Consistent discretization}

To make the calculations and illustrations simpler, we will
concentrate on three dimensional gravity, but the reader will readily
notice that there is no obstruction to applying the same reasonings in
$3+1$ dimensions. Given a simplicial approximation to a three
dimensional manifold, one can approximate the Einstein action (with a
cosmological term), as a sum over the edges (``hinges'') of the
decomposition plus a sum over the simplices\cite{Re},
\begin{equation}
S= k \sum_{h}  \ell_{h} \delta_{h} + \lambda
\sum_{\sigma } V_\sigma
\end{equation}
where the first sum is over all hinges and the second over 
all simplices, $\ell_{h}$ is the length of the hinge $h$ and 
$\delta_{h}$ is the
deficit angle around the hinge, i.e. $\delta_{h} = 2\pi - \sum_{\sigma_h}
\Theta(\sigma_h)$ where $\Theta(\sigma_h)$ is the angle formed
by the two faces of the simplices $\sigma_h$ that end in the hinge $h$.
$V_\sigma$ is the volume of the simplex $\sigma$
(in our three dimensional case, a tetrahedron). The constants $k$ and
$\lambda$ are related to Newton's constant and the cosmological
constant. A more explicit expression (see for instance David\cite{david})
can be given involving the values of the volumes of the two (in three
dimensions) faces which share the hinge $h$, $\sin \Theta(\sigma_h) =
{3/2} V_{\sigma_h} \ell_h /(A_{\sigma_h} A'_{\sigma_h})$, where
$A$ and $A'$ are the areas of the two triangles adjacent to $h$ in
the simplex $\sigma_h$. This in turn can be used to give an
expression that is purely a function of the lengths of the hinges,
using the Cayley--Menger determinants. We do not quote its 
explicit expression for brevity. 

In order to have a formulation that is amenable to a canonical
treatment that is uniform, in the sense that one has the same
treatment at all points on the lattice, one needs to make
certain assumptions about the regularity of the simplicial
decomposition chosen. This requirement can be somewhat relaxed and
our method still applies, but in a first approach we will 
consider a regular decomposition as shown in figure 1. We have
divided space-time in prisms (1 and 2 in the figure, for example),
and each prism in turn can be decomposed into three tetrahedra
(in the case of prism 2 the tetrahedra would be given by vertices $ABB'D$,
$AB'D'D$,$A'E'DD'$).
\begin{figure}[htbp]
  \centerline{
\includegraphics[height=3.3cm]{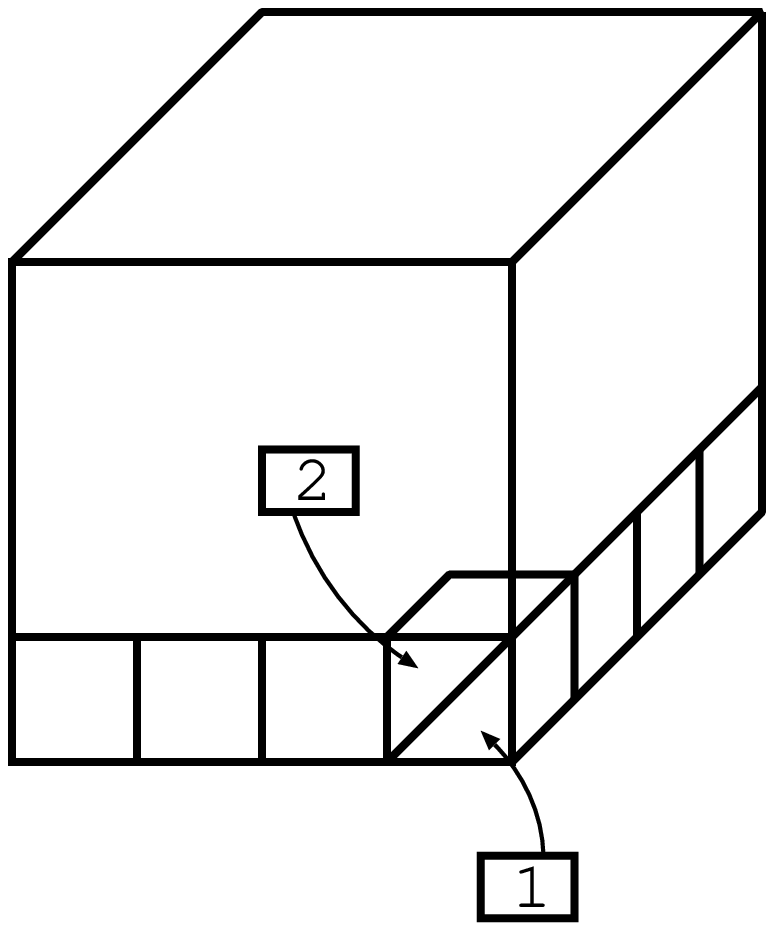}
\includegraphics[height=3.3cm]{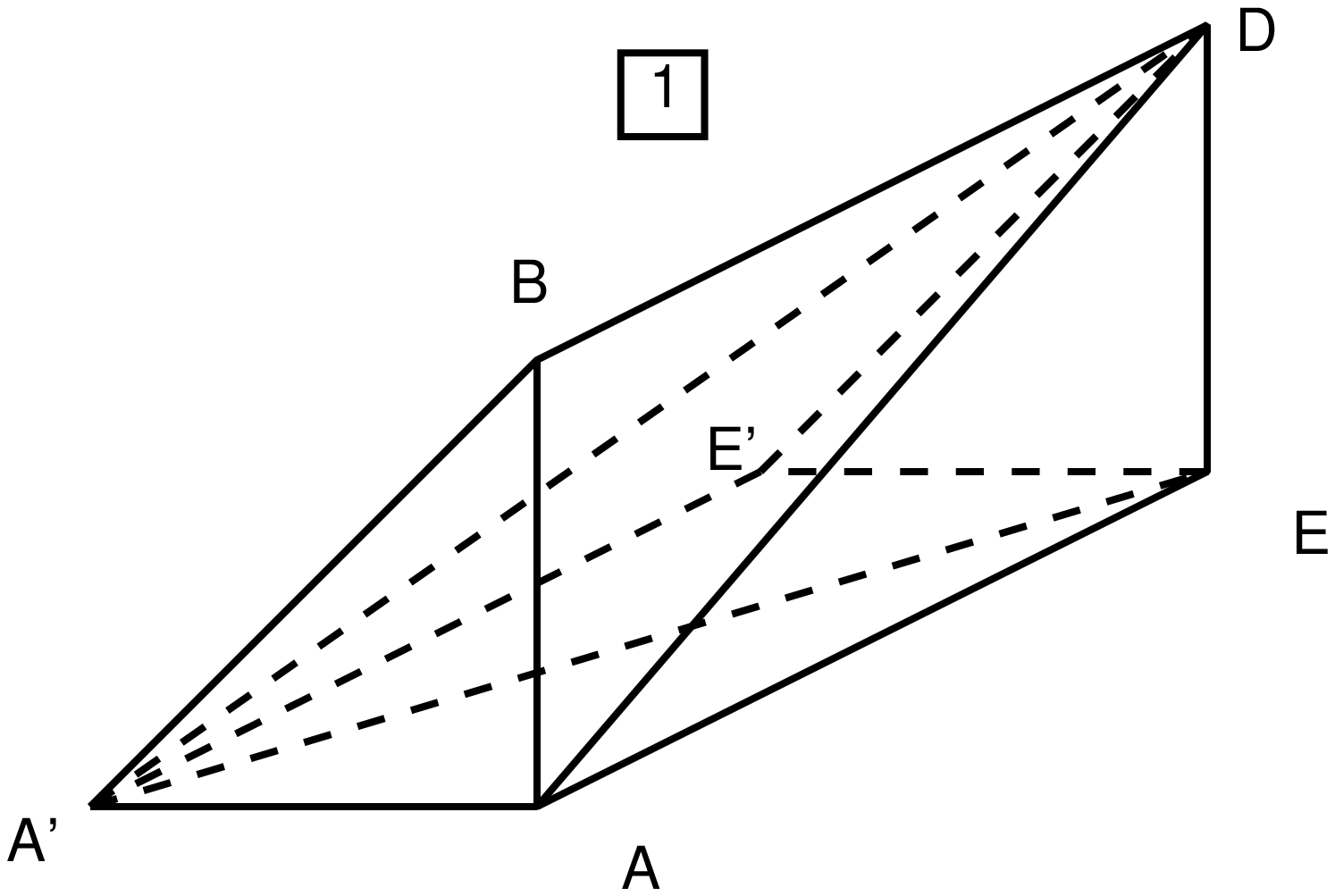}
\includegraphics[height=3.3cm]{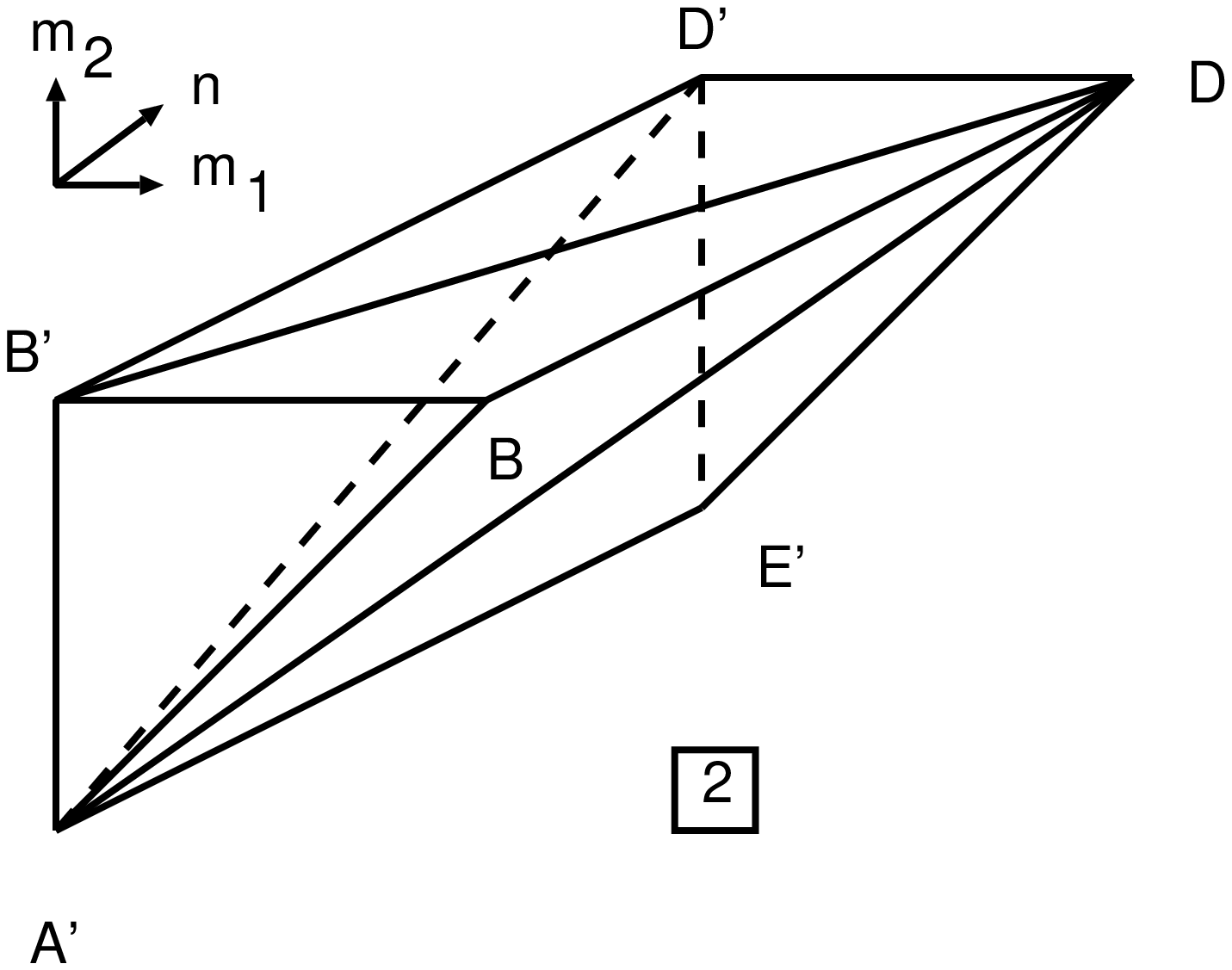}
}
\caption{The simplicial decomposition considered. The figures on the
right show prisms number 1 and 2 respectively, the other prisms are
obtained by reflection and periodicity. The hinge length variables
$\ell_i$ are assigned to the hinges in the following way:
$A'A\mapsto\ell_1$, $A'B\mapsto\ell_2$, $A'B'\mapsto\ell_3$,
$A'E\mapsto\ell_4$, $A'D\mapsto\ell_5$, $A'D'\mapsto\ell_6$
$A'E'\mapsto\ell_7$.}
\end{figure}

To construct a Lagrangian picture for the previous action we consider
two generic ``instants of time'' $n$ and $n+1$, as indicated by the
direction labeled $n$ in figure 1. We wish to construct an action of
the form $S =\sum_n L(n,n+1)$ where the Lagrangian $L(n,n+1)$ depends
on variables only at instants $n$ and $n+1$. We choose one of the
fundamental cubes (union of prisms 1 and 2 in the figure), choose a
conventional vertex in the cube labeled by $n,m_1,m_2$ in the
lattice. Notice that the use of the cubes is just for convenience, the
framework is based on prisms that have a triangular spatial basis and
therefore can tile any bidimensional spatial manifold. The variables
we will consider are the lengths $\ell_1,\ldots,\ell_7$ emanating from
the vertex, as designated in the figure. A similar construction is
repeated for each fundamental cube. The Lagrangian that reproduces the
Regge action is given by a function
\begin{eqnarray}
L(n,n+1)&=&\sum_{m_1,m_2}L\left(\frac{}{}\ell_1(n,m_1,m_2),\ldots,
\ell_7(n,m_1,m_2),\right.\\
&&\left.\ell_1(n+1,m_1,m_2),\ell_2(n+1,m_1,m_2),
\ell_3(n+1,m_1,m_2)\frac{}{}\right),\nonumber
\end{eqnarray}
that includes step functions that enforce the triangle inequalities
between the hinge length variables.

Up to now we have kept the discussion generic, but we should now make
things more precise, in dealing with either the Euclidean or the
Lorentzian case. In the former, all angles and quantities involved are
real. In the Lorentzian case, angles can become complex. Moreover
lengths can be time-like or space-like. Null intervals can also be
considered, but make the formulas more complicated, so for simplicity
we do not consider them here.  We will take all lengths as positive
numbers, irrespective of the space-like or time-like character of the
underlying hinge. In the above construction we have chosen the
decomposition in such a way that the hinge $\ell_7$ is time-like and
all other hinges are space-like. The formulas presented above (for the
angles, for instance) are valid in both the Euclidean and Lorentzian
case, but in the latter volumes, areas and length may have to be
considered as imaginary numbers. All volumes involving a time-like
direction are real, and in the construction these are the only ones
involved.  Areas are imaginary if they involve one time-like direction and
real if they do not. Lengths are real if they are
time-like and purely imaginary if they are space-like. With these
conventions, dihedral angles around time-like directions are real (for
instance around $\ell_7$), and dihedral angles around space-like
directions are complex. Some can be purely imaginary (for instance
rotation around $A'A$ in tetrahedron $AA'BD$) which correspond to
Lorentz boosts, or complex (for instance rotation around $AB$ in the
aforementioned tetrahedron) which does not correspond to a Lorentz
transformation (it traverses the light cone). There is one further
point to consider. In the expression for the deficit angle the term
$2\pi$ is present for hinges that span from the base of the elementary
cube to the top cover. For hinges lying entirely within the base or
the top cover the term is $\pi$. With these conventions the Lagrangian
$L(n,n+1)$ turns out to be real and the sum yields the correct action
avoiding over counting. For a more detailed discussion of angles in
the Lorentzian case see Sorkin\cite{sorkin}.

We now proceed to treat this action with the ``consistent
discretization'' approach. We consider as configuration variables 
$\ell_1,\ldots,\ell_7$ and define their canonical momenta,
\begin{eqnarray}
P_{\ell_i}(n+1) &=& {\partial L(n,n+1) \over \partial \ell_i(n+1)},\\
P_{\ell_i}(n) &=& -{\partial L(n,n+1) \over \partial \ell_i(n)}.
\end{eqnarray}

Here one is faced with several constraints. Notice that variables
$\ell_4,\ldots,\ell_7$ are ``Lagrange multipliers'' since the
Lagrangian does not depend on their value at instant $n+1$ and
therefore their canonical momenta vanish. The $P_{\ell_1},\ldots,
P_{\ell_3}$ only depend on links at level $n$ and therefore are
constraints among the variables. The system of equations determines
variables $\ell_4,\ldots,\ell_7$ and their momenta in terms of the
other variables so they can be eliminated.  The resulting canonical
pairs are $\ell_1,\ldots,\ell_3$ $P_{\ell_1},\ldots, P_{\ell_3}$. The
remaining equations are evolution equations for these variables and
there are no constraints left (in the sense of dynamical constraints,
the variables are still constrained by the usual triangle
inequalities). The evolution equations are a true canonical
transformation from the variables at instant $n$ to the variables at
instant $n+1$. This canonical transformation has as generating
function $-L(n,n+1)$, viewed as a type 1 canonical transformation,
where in the Lagrangian the variables $\ell_4,\ldots,\ell_7$ have been
replaced via the equations that determine them. The reader unfamiliar
with the ``consistent discretization'' approach may question the
legitimacy of this procedure in the sense of yielding a true canonical
structure, however it was discussed\cite{dirac} how the canonical
structure arises in detail through a generalization of the Dirac
procedure for discrete systems.

This concludes the classical discussion. We have reduced the 
Regge formulation to a well defined, unconstrained canonical system
where the discrete time evolution is implemented as a canonical 
transformation. Some of the original dynamical variables are
eliminated from the formulation using the equations of motion. In the
usual ``consistent discretizations'' the variables that are eliminated
are the Lagrange multipliers. Here the links that get determined 
can be viewed as playing a similar role. The equations that 
determine these variables are a complicated non-linear system. 
As in the usual ``consistent discretization'' approach, one may
have concerns that the solutions of the non-linear system could 
fail to be real, or could become unstable. We now have experience
with consistent discretizations of mechanical systems and of 
Gowdy cosmologies, which have field theoretic degrees of freedom 
and the evidence suggests that one can approximate the continuum
theory well in spite of potential complex solutions and multi-valued
branches\cite{Gowdy}. We expect a similar picture to occur in 
Regge calculus.

\section{Quantization}

Turning our attention to quantization, as usual in the ``consistent
discretization'' approach, the hard conceptual issues are sidestepped
since the theory is constraint-free. The task at hand is to 
implement the canonical transformation that yields the discrete
time evolution as a unitary quantum operator that implements
the discrete classical equations of motion as quantum operatorial
equations. This will in a generic situation be computationally 
intensive, but conceptually clear. It should be noted that the
resulting unitary operator differs significantly from the ones
that have been historically proposed in path integral approaches
based on Regge calculus. The usual approach to a path integral
would be to compute,
\begin{eqnarray}
&&\int \Pi_{i=1,7,n,\vec{m}} d\ell_i(n,\vec{m})
\mu(\ell_1(n,\vec{m}),\ldots,\ell_7(n,\vec{m})) \times\\
&&\times \exp\left(i\sum_{n',\vec{m'}} L(\ell_1(n,\vec{m}),\ldots,\ell_7(n,\vec{m})),
\ell_1(n+1,\vec{m}),\ldots,\ell_3(n+1,\vec{m})\right),\nonumber
\end{eqnarray}
with $\mu$ a measure that presumably should enforce the constraints
of the theory. On the other hand, in our approach one would have
something like
\begin{eqnarray}
&&\int \Pi_{i=1,3,n,\vec{m}} d\ell_1(n,\vec{m})
\mu(\ell_1(n,\vec{m}),\ldots,\ell_3(n,\vec{m}))\times \\
&&\times\exp\left(i\sum_{n',\vec{m'}} L'(\ell_1(n,\vec{m}),\ldots,\ell_3(n,\vec{m}),
\ell_1(n+1,\vec{m}),\ldots,\ell_3(n+1,\vec{m})))\right),\nonumber
\end{eqnarray}
where $L'$ is obtained by substituting in $L$ the values of the
``Lagrange multipliers'' $\ell_{4},\ldots,\ell_7$
obtained from their equations of motion. $\mu$ is uniquely determined when one determines the 
unitary transformation that implements the dynamics (examples
of this in cosmological situations can be seen in our paper\cite{cosmo}). So we
see that we have eliminated some of the variables and the constraints
of the theory and the path integral is uniquely defined by the
consistent discretization approach.

\section{Discussion}

Some concerns might be raised about the limitations imposed on our
framework by the choice of initial lattice. We have chosen to use a
lattice that is topologically cubical. This sets a well defined
framework in which to construct a Lagrangian evolution between two
spatial hypersurfaces. The cubic lattice is not a strict
requirement. It would be enough to have two ``close in time''
space-like hypersurfaces with the same simplicial decomposition in both
for us to be able to set up our framework and start evolving. This can
encompass quite a range of geometrical situations. It is however,
inevitable that one should give up some arbitrariness in the
space-time simplicial decomposition if one wishes to have a canonical
structure. It is interesting that the structure imposed is such
that it automatically eliminates the ``spikes'' (thin simplices
arbitrarily large in the time-like direction).

\begin{figure}[htbp]
  \centerline{
\includegraphics[height=4cm]{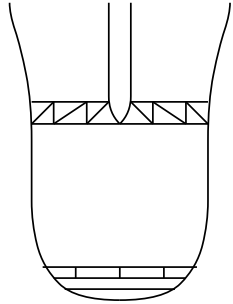}
}
\caption{The framework can handle topology change.}
\label{figure1}
\end{figure}

It also worthwhile emphasizing that the framework can, with 
relatively simple additions, incorporate topology change. The idea
is depicted in figure 2. There we see a point where there is topology
change where the legs of the pair of pants separate. For that to 
happen one would need to modify at the hypersurface the explicit
form of the Lagrangian $L(n,n+1)$. It is interesting that from there
on one can continue without further altering the framework and that
at all times the number of variables involved has not changed. The 
picture also shows how one would handle an initial ``no boundary'' 
type singularity. Here one would have to ``by hand'' add links as
the time evolution progresses forward. These variables are free
as long as one does not wish to match some final end state for the
evolution. If one has, however, specified initial and final data 
for the evolution, one finds that constraints appear that determine
the values of the lengths of the extra links added.

As an example of the framework, one can work out explicitly the
evolution of a a $2+1$ dimensional space-time consistent of a four
adjacent ``unit cubes'' of the type we have considered with fixed
outer boundary conditions. In this case the initial data consists of
eight lengths, the other initial lengths are determined by the
boundary conditions. As we discussed, having the data at level $n$ and
$n+1$ one can determine the ``Lagrange multiplier'' links that have to
be substituted in the Lagrangian to generate the canonical
transformation between the initial and final data.  This
transformation is later to be implemented unitarily upon quantization.
In our formalism it can happen that the ``Lagrange multipliers'' are
not entirely determined by this procedure (for other examples where
this happens, including BF theory, see our paper\cite{discrete}). In such case
the resulting theory has true constraints and true Lagrange
multipliers. In this example this happens.  One is finally left with a
canonical transformation dependent on 3 parameters (this result is
also true if one considers an $N\times N$ adjacent unit cube system).
The presence of free parameters also requires modifications in the
path-integral formulas listed above as well. It should be emphasized
that the equations determining the Lagrange multipliers, even in this
simplified case, are complicated coupled non-linear equations that
have a complexity not unlike those in $3+1$ dimensions. What makes
them easy to solve is the knowledge that the Regge equations of motion
correspond in this case to flat space-time.  The canonical
transformation can be implemented unitarily and the quantization
completed.  We will discuss the example in detail in a separate
publication.

\section{Conclusions}

We have applied the ``consistent discretization'' approach to Regge
calculus. We see that it leads to a well defined constraint-free
canonical formulation, that is well suited for quantization. The
approach can incorporate topology change.  Although we have limited
the equations to the three dimensional case for simplicity, we have
never used any of the special properties of three dimensional gravity
and it is clear that the construction can be carried out in an
arbitrary number of dimensions.  It is interesting to notice that one
of the original motivations for the construction of the ``consistent
discretization'' approach was the observation by Friedman and 
Jack\cite{FrJa} that in canonical Regge calculus the Lagrange multipliers
failed to be free. It can be viewed as if this point of view has now
been exploited to its fullest potential, offering a well defined
computational avenue to handle classical and quantum gravity.
\section*{Acknowledgments}

We dedicate this paper to Rafael Sorkin on the occassion of his 60th
birthday.  We wish to thank Luis Lehner and Jos\'e A. Zapata for
comments.  This work was supported by grant NSF-PHY0244335,
NASA-NAG5-13430 and funds from the Horace Hearne Jr. Laboratory for
Theoretical Physics and CCT-LSU.

\end{document}